\documentclass{iopart}
\usepackage{graphicx}  
\usepackage{latexsym}
\usepackage{amssymb}

\jl{6}        
\eqnobysec    

\def\beq{\begin{equation}}
\def\eeq{\end{equation}}

\def\rmd{{\rm d}}

\def\version{\today}

\let\Tilde=~

\begin{document}

\begin{flushright}
Current version: \version \\
\end{flushright}

\title
[Relativistic Poynting-Robertson effect]
{The general relativistic Poynting-Robertson effect}

\author{
Donato Bini${}^*{}^\S$,
Robert T. Jantzen${}^\P{}^\S$ and 
Luigi Stella$^\dag$
}
\address{
  ${}^*$\
Istituto per le Applicazioni del Calcolo ``M. Picone'', CNR I-00161 Rome, Italy
}
\address{
  ${}^\S$\
  International Center for Relativistic Astrophysics - I.C.R.A.,
  University of Rome ``La Sapienza'', I-00185 Rome, Italy
}
\address{
  ${}^\P$\
Department of Mathematical Sciences, Villanova University, Villanova, PA 19085,  USA
}
\address{
  ${}^\dag$\
Osservatorio Astronomico di Roma, via Frascati 33, I-00040 Monteporzio Catone (Roma), Italy
}

\begin{abstract}
The general relativistic version is developed for Robertson's discussion of the Poynting-Robertson effect that he based on special relativity and Newtonian gravity for point radiation sources like stars. The general relativistic model uses a test radiation field of photons in outward radial motion with zero angular momentum in the equatorial plane of the exterior Schwarzschild or Kerr spacetime.
\end{abstract}

\pacno{04.20.Cv}

\section{Introduction}

For a small body like a grain of dust orbiting a star, the radiation pressure of the light emitted by the star in addition to the direct effect of the outward radial force
exerts a drag force on the body's motion which causes it to fall into the star unless the body is so small that the radiation pressure pushes it away from the star. Called the Poynting-Robertson effect, it was first investigated by J.H. Poynting in 1903 \cite{poynting} using nonrelativistic physics and Newtonian gravity and then later re-calculated in 1937 
using special relativity and Newtonian gravity by H.P. Robertson \cite{robertson},
who also stated the leading general relativistic correction to his slow motion calculation, namely the perihelion precession for the quasi-Newtonian orbits. 
These calculations were revisited by Wyatt and Whipple in 1950 \cite{wyatt} for applications to meteor orbits, making more explicit Robertson's calculations for slowly evolving elliptical orbits and slightly extending them. 

The drag force is easily \emph{naively} understood as an aberration effect \cite{wiki}: if the body is in a circular orbit, for example, the radiation pressure is radially outward from the star, but in the rest frame of the body, the radiation appears to be coming from a direction slightly towards its own direction of motion, and hence a backwards component of force is exerted on the body which acts as a drag force. If the drag force dominates the outward radial force, the body falls into the star.
For the case in which a body is momentarily at rest, a critical luminosity similar to the Eddington limit for a star \cite{eddington}
exists at which the inward gravitational force balances the outward radiation force, a critical value separating radial infall from radial escape. Similarly for a body initially in a circular orbit, there are two kinds of solutions: those in which the body spirals inward or spirals outward, depending on the strength of the radiation pressure.

We now consider this problem in the context of a test body in orbit in a spherically symmetric Schwarzschild spacetime without the restriction of slow motion, and then in the larger context of an axially symmetric Kerr spacetime while developing the equations for a more general stationary axially symmetric spacetime. The finite size of the radiating body is ignored. 
Guess took this into account to extend Robertson's calculations in 1962 \cite{gues}, and Abramowicz, Ellis and Lanza generalized this for purely radial motion in the exterior Schwarzschild spacetime to model jets and solar winds, using numerical and qualitative methods to classify the trajectories of the test body into seven different categories \cite{abr-ell-lan}, introducing the key idea of a ``saturation velocity" for which the test body remains momentarily unaccelerated due to the balancing of the gravitational and radiation forces. Miller and Lamb then considered the case of arbitrary motion around a slowly rotating star \cite{lam-mil,mil-lam3}, after reconsidering the nonrotating case \cite{mil-lam1}.

In order to see the key features of the general relativistic case without the additional complications required to take into account the finite size of the radiating body,
we return to the simple Robertson scenario and look only at how strong gravitational fields and arbitrary motion affect the problem. The photon flux from the central body is modeled by test photons in outward radial motion with respect to the locally nonrotating observers, namely photons with vanishing conserved angular momentum. While developing the basic equations for 
planar motion of both the test body and test photons in a stationary axisymmetric spacetime, we consider their solution only for motion in the equatorial plane of the Schwarzschild and Kerr spacetimes. Of course it only makes sense to consider the exterior solutions for radii larger than some minimum radius $R$ outside the horizon in order to model the geometry outside a star (or some other physical source) of radius $R$ producing the outflow of radiation. 

\section{Stationary, axisymmetric and reflection-symmetric spacetimes} 

Using a Boyer-Lindquist-like coordinate system $\{t,r,\theta,\phi \}$ adapted to the spacetime symmetries, i.e., with $\partial_t$ (timelike) and $\partial_\phi$ (spacelike, with closed coordinate lines) a pair of commuting Killing vectors, the metric of a stationary axisymmetric spacetime can be expressed by a line element of the form
\beq
\label{metr_gen}
\rmd s^2=    g_{tt} \rmd t^2 
         + 2 g_{t\phi} \rmd t \rmd \phi 
         + g_{\phi\phi}\rmd \phi^2 
         + g_{rr}\rmd r^2 
         + g_{\theta\theta} \rmd \theta^2\,,
\eeq
where all the metric coefficients depend only on $r$ and $\theta$, provided that the metric belongs to the most interesting class of orthogonally symmetric such metrics \cite{ES}. 
As is the case for the familiar black hole spacetimes, we further require the metric to be reflection-symmetric with respect to the equatorial plane $\theta=\pi/2$, within the allowed range $\theta\in[0,\pi]$ of this angular coordinate, where $(r,\phi)$ then behave like flat space polar coordinates and will be treated as such to display orbit plots below. Our calculations will be limited to this plane for simplicity.

The time coordinate lines, when timelike, are the world lines of the static observers. 
The zero angular momentum observer (ZAMO) family of fiducial observers has instead a 4-velocity $n$ characterized as that normalized linear combination of the two given Killing vectors which is orthogonal to $\partial_\phi$ and future-pointing, and it is the unit normal to the time coordinate hypersurfaces
\beq
\label{n}
n=N^{-1}(\partial_t-N^{\phi}\partial_\phi)\,,
\eeq
where $N=(-g^{tt})^{-1/2}$ and $N^{\phi}=g_{t\phi}/g_{\phi\phi}$ are the lapse function and only nonvanishing component of the shift vector field respectively. 
Our discussion is limited to those regions of spacetime where the time coordinate hypersurfaces are spacelike: $g^{tt}<0$. 
A suitable orthonormal frame adapted to the ZAMOs and invariant under the symmetry group action is given by
\beq
\label{zamoframe}
e_{\hat t}=n\,,\quad
e_{\hat r}=\frac1{\sqrt{g_{rr}}}\partial_r\,,\quad
e_{\hat \theta}=\frac1{\sqrt{g_{\theta \theta }}}\partial_\theta\,,\quad
e_{\hat \phi}=\frac1{\sqrt{g_{\phi \phi }}}\partial_\phi\,,
\eeq
with dual 
\beq
\fl\quad
\omega^{{\hat t}}
  = N \rmd t\,,\quad \omega^{{\hat r}}
  = \sqrt{g_{rr}} \,\rmd r\,,\quad 
\omega^{{\hat \theta}}
  = \sqrt{g_{\theta \theta }} \,\rmd \theta\,,\quad
\omega^{{\hat \phi}}
  = \sqrt{g_{\phi \phi }}(\rmd \phi+N^{\phi}\rmd t)\,,
\eeq
so that the line element (\ref{metr_gen}) can be also expressed in the form
\beq
\rmd s^2 = -N^2\rmd t^2 +g_{\phi \phi }(\rmd \phi+N^{\phi}\rmd t)^2 + g_{rr}\rmd r^2 +g_{\theta \theta}\rmd \theta^2\,. 
\eeq

The accelerated ZAMOs are locally nonrotating in the sense that their vorticity vector $\omega(n)$ vanishes, but they have a nonzero expansion tensor $\theta(n)$ whose nonzero
components can be completely described by the shear vector
$\theta_{\hat \phi}(n)^\alpha = \theta(n)^\alpha{}_\beta\,{e_{\hat\phi}}^\beta$, namely
\beq
\label{exp_zamo}
\theta(n) = e_{\hat\phi}\otimes\theta_{\hat\phi}(n)
           +\theta_{\hat\phi}(n)\otimes e_{\hat\phi}\,.
\eeq 
Since the expansion scalar $\theta(n)^\alpha{}_\alpha$ is zero, the expansion and shear tensors coincide.

The nonzero ZAMO kinematical quantities (acceleration $a(n)=\nabla_n n$ and shear tensor) and the conveniently defined Lie relative curvature vector \cite{idcf2,bjdf} only have nonzero components in the $r$-$\theta$ 2-plane of the tangent space
\begin{eqnarray}
\label{accexp}
\fl\quad
a(n) & = & a(n)^{\hat r} e_{\hat r} + a(n)^{\hat\theta} e_{\hat\theta}
 =\partial_{\hat r}(\ln N) e_{\hat r} + \partial_{\hat\theta}(\ln N)  e_{\hat\theta}
\,,
\nonumber\\
\fl\quad
\theta_{\hat\phi}(n) 
& = & \theta_{\hat\phi}(n)^{\hat r} e_{\hat r} + \theta_{\hat\phi}(n)^{\hat\theta} e_{\hat \theta} 
  = -\frac{\sqrt{g_{\phi\phi}}}{2N}\,[\partial_{\hat r} N^\phi e_{\hat r} + \partial_{\hat\theta} N^\phi e_{\hat \theta}
\,, 
\nonumber\\
\fl\quad
k_{(\rm lie)}(n)
& = & k_{(\rm lie)}(n)_{\hat r} e_{\hat r} + k_{(\rm lie)}(n)_{\hat\theta} e_{\hat\theta}
 = -[\partial_{\hat r}(\ln \sqrt{g_{\phi\phi}}) e_{\hat r} + \partial_{\hat\theta}(\ln \sqrt{g_{\phi\phi}})e_{\hat\theta}]
\,.
\end{eqnarray}
Here $\partial_{\hat r}\equiv e_{\hat r}$ and $\partial_{\hat \theta}\equiv e_{\hat \theta}$.
In the static limit $N^\phi\to0$, the shear vector $\theta_{\hat\phi}(n)$ vanishes.

Let a pure electromagnetic radiation field be superposed as a test field on the gravitational background described by the metric (\ref{metr_gen}),  with the energy-momentum tensor
\beq
\label{ten_imp}
T^{\alpha\beta}=\Phi^2 k^\alpha k^\beta, \qquad k^\alpha k_\alpha=0\,,
\eeq
where $k$ is assumed to be tangent to an  affinely parametrized outgoing null geodesic in the equatorial plane, i.e., $k^\alpha \nabla_\alpha k^\beta=0$ with $k^\theta=0$. 
We will only consider photons in the equatorial plane which are
in outward radial motion with respect to the ZAMOs, namely with 4-momentum
\beq
\label{eq:phot}
k= E(n)[n+\hat \nu(k,n)], \qquad  
\hat \nu(k,n)=e_{\hat r}\,,
\eeq
where $E(n)= E/N$ is the relative energy of the photon and $E=-k_t$ is the conserved energy associated with the timelike Killing vector field and $L=k_\phi=0$ is the vanishing conserved angular momentum associated with the rotational Killing vector field, while
$\hat \nu(k,n)$ defines the unit vector direction of the relative velocity.
For the Schwarzschild case, these orbits are radial geodesics with respect to the static observers tied to the coordinate system, but for the Kerr case, they are dragged azimuthally by the rotation of the spacetime with respect to the coordinates.

Since $k$ is completely determined, the coordinate dependence of the quantity $\Phi$  then follows 
from the conservation equations  $T^{\alpha\beta}{}_{;\beta}=0$, and will only depend on $r$ in the equatorial plane due to the axial symmetry.
From Eq.~(\ref{ten_imp}) using the geodesic condition for $k$, these can also be written as
\beq
\label{flux_cons}
\nabla_\beta (\Phi^2 k^\beta)=0\,,\qquad 
k^\beta \partial_\beta \Phi^2 + k^\beta{}_{;\beta}\Phi^2=0\,.
\eeq
Integrating this equation requires an expression for the divergence of $k$, namely
\beq
\fl\quad
k^\alpha{}_{;\alpha}=
E(n) \frac{\partial}{\partial \hat r}\ln [N E(n) \sqrt{g_{\theta\theta}g_{\phi\phi}}]=
\frac{E}{N}\frac{\partial}{\partial \hat r}\ln [\sqrt{g_{\theta\theta}g_{\phi\phi}}]\,.
\eeq
$\Phi$ is then easily determined taking into account that $k^\beta \partial_\beta=k^{\hat r}\partial_{\hat r}=E/N\partial_{\hat r}$, so
\beq
\Phi =  (g_{\theta\theta}g_{\phi\phi})^{-1/4} \Phi_0 \,,
\eeq
as we will see
explicitly below for the Schwarzschild and Kerr cases.

Consider now a test particle of mass $m$ moving in the equatorial plane $\theta=\pi/2$ accelerated by its interaction with the radiation field, i.e., with 4-velocity 
\beq\label{polarnu}
\fl\quad 
U=\gamma(U,n) [n+ \nu(U,n)]\,,\quad 
\nu(U,n)\equiv \nu^{\hat r}e_{\hat r}+\nu^{\hat \phi}e_{\hat \phi} 
=  \nu \sin \alpha e_{\hat r}+ \nu \cos \alpha e_{\hat \phi}  \,,
\eeq
where $\gamma(U,n)= (1-||\nu(U,n)||^2)^{-1/2}$ is the Lorentz factor and the abbreviated notation $\nu^{\hat a}=\nu(U,n)^{\hat a}$ has been used. In a similarly abbreviated notation, $\nu = ||\nu(U,n)||$ and $\alpha$ are the magnitude of the spatial velocity $\nu(U,n)$ and its
polar angle  measured clockwise from the positive $\phi$ direction in the $r$-$\phi$ tangent plane, while $\hat\nu=\hat \nu(U,n)$ is the associated unit vector.
Note that $\alpha=0$ corresponds to azimuthal motion with respect to the ZAMOs (i.e., in the $\phi$ direction only), while 
$\alpha=\pm \pi/2$ corresponds to (outward/inward) radial motion with respect to the ZAMOs.

Using the expression (\ref{n}) for $n$ leads to the coordinate components of $U$
\begin{eqnarray}
\label{Ucoord_comp}
&& U^t\equiv \frac{\rmd t}{\rmd \tau}=\frac{\gamma}{N}, \qquad U^r\equiv \frac{\rmd r}{\rmd \tau}
=\frac{\gamma \nu^{\hat r}}{\sqrt{g_{rr}}}\,, 
\nonumber \\
&& U^\theta\equiv \frac{\rmd \theta}{\rmd \tau}=0, \qquad U^\phi\equiv \frac{\rmd \phi}{\rmd \tau}=
\frac{\gamma \nu^{\hat \phi}}{\sqrt{g_{\phi\phi}}} 
-\frac{\gamma N^\phi}{N}\,,
\end{eqnarray}
where $\tau$ is the proper time parameter along $U$. Solving these for the magnitude and polar angle leads to
\begin{eqnarray}
\tan \alpha &=&\sqrt{ 
\frac{g_{rr}}{g_{\phi\phi}}
}\, 
\frac{\rmd r}{\rmd t}\left( \frac{\rmd \phi}{\rmd t}  +N^\phi\right)^{-1}\,, 
\nonumber \\
\nu&=& 
\frac{1}{N}\sqrt{g_{rr} \left( \frac{\rmd r}{\rmd t} \right)^2
         + g_{\phi\phi} \left( \frac{\rmd \phi}{\rmd t}+N^\phi \right)^2} \,.
\end{eqnarray}

Scattering of radiation as well as the momentum-transfer cross section $\sigma$ (assumed to be a constant) of the particle is independent of the direction and frequency of the radiation; therefore the associated force is given by 
\beq
{\mathcal F}_{\rm (rad)}(U)^\alpha = -\sigma P(U)^\alpha{}_\beta \, T^{\beta}{}_\mu \, U^\mu
\eeq
where $P(U)^\alpha{}_\beta=\delta^\alpha_\beta+U^\alpha U_\beta$ projects orthogonally to $U$.
The equation of motion of the particle then becomes
\beq
\label{eq:motog}
m a(U) = {\mathcal F}_{\rm (rad)}(U)\,,
\eeq
where $m$ is the mass of the particle and $a(U)=\nabla_U U$ is its 4-acceleration.

To examine the content of this equation, in addition to the decomposition (\ref{eq:phot}) with respect to the ZAMOs, 
it is convenient to decompose $k$ with respect to $U$ as well
\beq
\label{diff_obg}
k = E(U)[U+\hat {\mathcal V}(k,U)]
 \equiv E(n)[n+\hat \nu(k,n)]\,.
\eeq
It then follows that
\beq
P(U)k=E(U)\hat {\mathcal V}(k,U)\,,\quad 
U\cdot k=-E(U)
\eeq
so that
\begin{eqnarray}
\fl\quad
{\mathcal F}_{\rm (rad)}(U)^\alpha&=&-\sigma \Phi^2 (P(U)^\alpha{}_\beta k^\beta)\, (k_\mu U^\mu)=
\sigma \, [\Phi E(U)]^2\, \hat {\mathcal V}(k,U)^\alpha\,.
\end{eqnarray}
Therefore we find
\beq
||a(U)||=\tilde \sigma [\Phi E(U)]^2\,,\qquad  
a(U)/||a(U)|| = \hat {\mathcal V}(k,U)\,,
\eeq
where $\tilde \sigma=\sigma/m$. 
Hereafter we will employ the simplified notation
\beq
\fl\qquad
||\nu(U,n)||=\nu\,, \quad 
\gamma(U,n) =\gamma\,,\quad 
\hat {\mathcal V}(k,U)=\hat {\mathcal V} \,.
\eeq

From Eq.~(\ref{diff_obg}), after scalar multiplication by $U$, one finds the relation 
\beq
\label{photon_en}
E(U)=\gamma E(n)[1-\nu(U,n)\cdot \hat \nu(k,n)]
\eeq
which leads to the following expression for $\hat {\mathcal V}$
\beq
\hat {\mathcal V}=\left[\frac{E(n)}{E(U)}-\gamma \right]n + \frac{E(n)}{E(U)}\hat \nu(k,n)-\gamma \nu(U,n)\,.
\eeq

Introducing the polar decomposition (\ref{polarnu}) of the particle velocity Eq.~(\ref{photon_en}) becomes
\beq
E(U)=\gamma E(n)[1-\nu\cos(\alpha-\pi/2)]
=\gamma E(n)[1-\nu\sin\alpha]\,,
\eeq 
so that
the frame components  of $\hat {\mathcal V}=\hat {\mathcal V}{}^{\hat t}n+\hat {\mathcal V}{}^{\hat r}e_{\hat r}+\hat {\mathcal V}{}^{\hat \phi}e_{\hat \phi}$ are given by
\beq
\label{eq:hatVu}
\fl\quad
\hat {\mathcal V}{}^{\hat t}=\gamma \nu \frac{\sin \alpha-\nu}{1-\nu\sin \alpha}\,,\quad 
\hat {\mathcal V}{}^{\hat r}=\frac{1}{\gamma (1-\nu \sin \alpha)} -\gamma \nu \sin\alpha\,,\quad 
\hat {\mathcal V}{}^{\hat \phi}= -\gamma \nu \cos\alpha\,.
\eeq
Note the relations
\begin{eqnarray}
\label{ort_VUaU}
\hat {\mathcal V}{}^{\hat t}
&=&\nu (\hat{\mathcal V}{}^{\hat r}\sin\alpha
          + \hat{\mathcal V}{}^{\hat \phi}\nu \cos\alpha)\,,\nonumber\\
a(U)^{\hat t}
&=& \nu [a(U)^{\hat r}\sin\alpha + a(U)^{\hat \phi}\nu \cos\alpha ]
\end{eqnarray}
which follow from the orthogonality of the pairs $(\hat{\mathcal V},U)$ and $(a(U),U)$.

Finally a straightforward calculation shows that the frame components of the 4-acceleration $a(U)$ in the equatorial plane (and hence the equations of motion) are given by
\begin{eqnarray}
\label{eq_fundam}
\fl
a(U)^{\hat t}
&=&   \gamma^2 \nu \sin \alpha 
      \left[a(n)^{\hat r}
         +2\nu\cos \alpha\, \theta(n)^{\hat r}{}_{\hat \phi}\right] 
         + \gamma^3\nu \frac{\rmd \nu}{\rmd \tau}
=\tilde\sigma [\Phi E(U)]^2 \hat {\mathcal V}^{\hat t}\,,\nonumber \\
\fl
a(U)^{\hat r}
&=& \gamma^2 [a(n)^{\hat r}+k_{\rm (lie)}(n)^{\hat r}\,\nu^2 \cos^2\alpha
            +2\nu\cos \alpha\, \theta(n)^{\hat r}{}_{\hat \phi}]\nonumber \\
\fl
&& + \gamma
 \left[\gamma^2 \sin\alpha \frac{\rmd \nu}{\rmd \tau} 
       +\nu \cos \alpha \frac{\rmd \alpha}{\rmd \tau}\right]
=\tilde\sigma [\Phi E(U)]^2 \hat {\mathcal V}^{\hat r}\, ,\nonumber \\
\fl
a(U)^{\hat \theta}&=& 0\, , \nonumber \\
\fl
a(U)^{\hat \phi}&=& -\gamma^2 \nu^2 \sin \alpha \cos \alpha\, k_{\rm (lie)}(n)^{\hat r} 
   + \gamma\left(
           \gamma^2 \cos \alpha  \frac{\rmd \nu}{\rmd \tau}-\nu\sin \alpha \frac{\rmd \alpha}{\rmd \tau}\right)\nonumber \\
\fl
&=& \tilde\sigma [\Phi E(U)]^2 \hat {\mathcal V}^{\hat \phi}\,.
\end{eqnarray}

The first of these equations is a linear combination (\ref{ort_VUaU}) of the remaining ones due to orthogonality of $a(U)$ and $U$. Using the expression for the components of ${\mathcal V}$
one obtains 
\begin{eqnarray}
\label{eqs:beta_eq_pi2_tau0}
\fl
\frac{\rmd \nu}{\rmd \tau}
&=& 
-\frac{\sin\alpha}{\gamma}[a(n)^{\hat r}+2\nu\cos \alpha\, \theta(n)^{\hat r}{}_{\hat \phi}]
  + \frac{A}{N^2 \sqrt{g_{\theta\theta}g_{\phi\phi}}}(1-\nu \sin \alpha)(\sin \alpha -\nu) \,,
\nonumber \\
\fl
\frac{\rmd \alpha}{\rmd \tau}
&=& 
 -\frac{\gamma\cos \alpha}{\nu}[a(n)^{\hat r}
 +2\nu\cos \alpha\, \theta(n)^{\hat r}{}_{\hat \phi}
 +\nu^2 k_{\rm (lie)}(n)^{\hat r}] 
\nonumber \\
\fl
&& + \frac{A}{N^2 \sqrt{g_{\theta\theta}g_{\phi\phi}}} \, \frac{(1-\nu \sin \alpha)\cos \alpha}{\nu} \,,
\nonumber\\
\fl 
\frac{\rmd r}{\rmd \tau}&=&
\frac{\gamma \nu\sin\alpha}{\sqrt{g_{rr}}}
\,,
\end{eqnarray}
where the positive constant $A$ is defined by
\beq
  A = \tilde\sigma \Phi_0^2 E^2
\eeq
in terms of which one has
\beq
    \tilde\sigma \Phi^2 E(n)^2 = \frac{A}{\sqrt{g_{\theta\theta} g_{\phi\phi}}\, N^2} \,.
\eeq

These equations admit the special solutions $\alpha=\pm\pi/2$ (radial outward/inward motion with respect to the ZAMOs) with
\beq
\frac{\rmd \nu}{\rmd \tau} 
= \pm\left[ -\frac{a(n)^{\hat r}}{\gamma} + \frac{A}{\sqrt{g_{\theta\theta} g_{\phi\phi}}\, N^2}(1\mp\nu)^2 \right]\,.
\eeq
Furthermore, if $\nu$ vanishes initially the condition that it remain zero is
\beq\label{eq:radialbalance}
   \frac{A}{\sqrt{g_{\theta\theta} g_{\phi\phi}}\, N^2} = a(n)^{\hat r}\,.
\eeq
This condition balancing the gravitational attraction and the radiation pressure gives the values $r_{\rm(crit)}$ of the radial coordinate where the particle comoves with the ZAMOs in a circular orbit. It will be discussed below when dealing with special cases. Recall that here $A$ contains a factor $1/m$ in $\tilde\sigma$, so that for a small enough $m$ one can always achieve this force balance at a given radius.

From the introduction it is clear that circular orbits cannot exist because of the drag force in the azimuthal direction.
A circular orbit would have constant $\alpha=0$ which requires using instead Eqs.~(\ref{eqs:beta_eq_pi2_tau0}), leading to
\beq
\label{eqs:beta_eq_pi2_tau_alpha}
\fl\quad
\frac{\rmd \nu}{\rmd \tau} = -\frac{A}{\sqrt{g_{\theta\theta} g_{\phi\phi}}\, N^2} \nu, \quad
0 = -\gamma[a(n)^{\hat r}+2\nu\, \theta_{\hat \phi}(n)^{\hat r}
+\nu^2 k_{\rm (lie)}(n)^{\hat r}]+\frac{A}{\sqrt{g_{\theta\theta} g_{\phi\phi}}\, N^2}\,.
\eeq
As we will see in detail below  when dealing with special cases, the quantity  $A/(\sqrt{g_{\theta\theta} g_{\phi\phi}} N^2)$ is a function of $r$ alone which in turn is a constant with respect to $\tau$ for a circular orbit. Therefore, from the second of Eqs.~(\ref{eqs:beta_eq_pi2_tau_alpha}), we would find a constant value for $\nu$ which is not compatible with the first of Eqs.~(\ref{eqs:beta_eq_pi2_tau_alpha}) unless $\nu=0$, which means $U=n$, i.e., the motion is comoving with the ZAMOs. The only surviving equation is associated with the radial component of the acceleration
reducing the situation to the previously discussed radial force balance. In fact if one chooses the speed so that the radial acceleration is initially zero, then the speed will decrease, causing the centripetal acceleration term to decrease so that the radial acceleration will become negative, and orbit will decay.

\section{Explicit spacetimes}

It is helpful to examine in sequence first the special relativistic effects involved in this drag force, then the addition of the gravitational field of a nonrotating source by considering the Schwarzschild spacetime, and then the addition of the source rotation finally in the context of the Kerr spacetime.
The equations of motion are perhaps best examined in detail directly in terms of the coordinate variables $t,r,\phi$. The nonzero
coordinate components of the four acceleration are
\begin{eqnarray}
\fl\quad
a(U)^t&=& 
\frac{\rmd^2 t }{\rmd \tau^2}+2\frac{\rmd t }{\rmd \tau}\frac{\rmd r }{\rmd \tau} a(n)_{\hat r}\sqrt{g_{rr}} 
+2\frac{\sqrt{g_{\phi\phi}}}{N}\frac{\rmd r }{\rmd \tau} \left[\left(N^\phi \frac{\rmd t }{\rmd \tau}+\frac{\rmd \phi }{\rmd \tau}\right)\sqrt{g_{rr}}\,
\theta_{\hat\phi}(n)^{\hat r}\right] \,,
\nonumber \\
\fl\quad 
a(U)^r&=& 
\frac{\rmd^2 r }{\rmd \tau^2}+(\partial_r \ln \sqrt{g_{rr}}) \left(\frac{\rmd r }{\rmd \tau}\right)^2 
+2\frac{\sqrt{g_{\phi\phi}}}{g_{rr}}\,\frac{\rmd t }{\rmd \tau} 
     \left(N^\phi \frac{\rmd t }{\rmd \tau}+\frac{\rmd \phi }{\rmd \tau}\right)N\theta_{\hat\phi}(n)^{\hat r}
\nonumber \\
\fl\quad 
&& 
+\frac{\sqrt{g_{\phi\phi}}}{g_{rr}} \, \left(N^\phi \frac{\rmd t }{\rmd \tau}+ \frac{\rmd \phi }{\rmd \tau} \right)^2 k_{(\rm lie)}(n)_{\hat r} 
+\frac{N^2}{\sqrt{g_{rr}}} \, a(n)_{\hat r}\left(\frac{\rmd t }{\rmd \tau}\right)^2 \,,
\nonumber \\
\fl\quad 
a(U)^\phi&=& 
\frac{\rmd^2 \phi }{\rmd \tau^2} -2\frac{\sqrt{g_{rr}}}{N\sqrt{g_{\phi\phi}}} \, \theta_{\hat\phi}(n)^{\hat r} \frac{\rmd r }{\rmd \tau}
\left[
 N^2 \left(\frac{\rmd t }{\rmd \tau}\right)^2
 + g_{\phi\phi}N^\phi \left(N^\phi \frac{\rmd t }{\rmd \tau}
 + \frac{\rmd \phi }{\rmd \tau} \right)
\right]
\nonumber \\ 
\fl\quad
&& 
-2N^\phi \sqrt{g_{rr}}\,a(n)_{\hat r}\frac{\rmd t }{\rmd \tau}\frac{\rmd r }{\rmd \tau} 
-2\sqrt{g_{rr}} \, k_{(\rm lie)}(n)_{\hat r} \frac{\rmd r }{\rmd \tau}
  \left(N^\phi \frac{\rmd t }{\rmd \tau}+\frac{\rmd \phi }{\rmd \tau}\right) 
\, . 
\end{eqnarray} 
In the case of a static spacetime where
$N^\phi=0$, $\theta_{\hat\phi}(n)^{\hat r}=0$, 
the above relations reduce to 
\begin{eqnarray}
\fl
a(U)^t&=& 
\frac{\rmd^2 t }{\rmd \tau^2}+2\frac{\rmd t }{\rmd \tau}\frac{\rmd r }{\rmd \tau} a(n)_{\hat r}\sqrt{g_{rr}} \,,
\nonumber \\
\fl
a(U)^r&=& 
 \frac{\rmd^2 r }{\rmd \tau^2}+(\partial_r \ln \sqrt{g_{rr}}) \left(\frac{\rmd r }{\rmd \tau}\right)^2
+\frac{\sqrt{g_{\phi\phi}}}{g_{rr}} \, \left(\frac{\rmd \phi }{\rmd \tau} \right)^2 k_{(\rm lie)}(n)_{\hat r} 
+\frac{N^2}{\sqrt{g_{rr}}} \, a(n)_{\hat r}\left(\frac{\rmd t }{\rmd \tau}\right)^2 ,
\nonumber \\ 
\fl
a(U)^\phi&=& 
\frac{\rmd^2 \phi }{\rmd \tau^2}  -2\sqrt{g_{rr}}\, k_{(\rm lie)}(n)_{\hat r} \,\frac{\rmd r }{\rmd \tau}\frac{\rmd \phi }{\rmd \tau} 
\,. 
\end{eqnarray} 

The nonvanishing coordinate components of the radiation force (per unit mass) in the general case are given by
\begin{eqnarray} 
\tilde {\mathcal F}_{\rm (rad)}(U)^t&=& \frac{Y}{N}\left[-1+N\left(\frac{\rmd t }{\rmd \tau}\right)X\right] \,,
\nonumber \\ 
\tilde {\mathcal F}_{\rm (rad)}(U)^r&=& \frac{Y}{\sqrt{g_{rr}}} 
\left[-1+\sqrt{g_{rr}}\left(\frac{\rmd r }{\rmd \tau}\right)X\right] \,,
\nonumber \\ 
\tilde {\mathcal F}_{\rm (rad)}(U)^\phi&=& \frac{Y}{N}\left[N^\phi+N\left(\frac{\rmd \phi }{\rmd \tau}\right)X\right]\,, 
\end{eqnarray} 
where 
$$ 
X=N \frac{\rmd t }{\rmd \tau}-\sqrt{g_{rr}}\frac{\rmd r }{\rmd \tau}, \quad Y=-\frac{AX}{N^2r\sqrt{g_{\phi\phi}}}\,. 
$$ 
In the static case only the $\phi$ component simplifies slightly due to $N^\phi=0$.

\subsection{Flat spacetime}

Consider first the flat spacetime case for
which the metric functions are
\beq
g_{tt}=-1\,, \quad 
g_{t\phi}=0\,, \quad 
g_{rr}=1\,, \quad 
g_{\theta\theta}=r^2\,, \quad 
g_{\phi\phi}=r^2\sin^2\theta\,,
\eeq
so that $N=1$ and $N^\phi=0$.
In this case, the ZAMOs are aligned with the coordinate time world lines and form a geodesic ($a(n)=0$) and nonexpanding ($\theta(n)=0$) congruence.
The Lie curvature vector associated with the $\phi$ coordinate lines only has a radial component with value $k_{\rm (lie)}(n)^{\hat r}=-1/r$.
Moreover, radially outgoing (geodesic) photons in the equatorial plane have 4-momentum
\beq
k=E(\partial_t + \partial_r)\,,
\eeq
Thus we find $E(n)=E$ and  $\Phi=\Phi_0/r$.

Therefore Eqs.~(\ref{eqs:beta_eq_pi2_tau0}) reduce to
\begin{eqnarray}
\frac{\rmd \nu}{\rmd \tau}
&=&  \frac{A(1-\nu \sin \alpha)(\sin \alpha -\nu)}{r^2} \,,\nonumber\\
\frac{\rmd \alpha}{\rmd \tau}
&=& -\nu\gamma\cos \alpha\, k_{\rm (lie)}(n)^{\hat r} 
    +\frac{A(1-\nu \sin \alpha)\cos \alpha}{r^2\nu}\,.
\end{eqnarray}
Furthermore Eqs.~(\ref{Ucoord_comp}) imply
\beq
\frac{\rmd r}{\rmd \tau}
=\gamma \nu \sin\alpha \,, 
\quad 
\frac{\rmd \phi}{\rmd \tau}
=\frac{\gamma \nu }{r}\cos\alpha\,,
\quad
\frac{\rmd r}{\rmd \phi}=r \tan\alpha\,.
\eeq

In the special case $\sin\alpha=\pm1$ of purely radial motion ($+$: outward, $-$: inward), one finds
\beq
\frac{\rmd \nu}{\rmd \tau}
=  \pm\frac{A(1\mp \nu)^2}{r^2} \,,\quad 
\frac{\rmd r}{\rmd \tau}
= \pm\gamma \nu \,,
\eeq
which leads to speeding up the outward motion and braking the inward motion as expected.
This orbit is described by
\beq
\frac{\rmd \nu}{\rmd r}
= \frac{A (1\mp \nu)^2}{ r^2 \nu \gamma}
\,,
\eeq
which can be easily integrated to yield a cubic relation between $\nu(r)$ and $r$.

Consider the behavior of the speed for purely circular initial motion $\sin\alpha=0$ where 
\beq
\frac{\rmd \nu}{\rmd \tau}
=  -\frac{A\nu}{r^2} \equiv -\tilde F_{\rm(drag)}
\eeq
gives the magnitude of the tangential drag force per unit mass on the orbit. If one takes the nonrelativistic Keplerian speed $\nu=(M/r)^{1/2}\ll 1$
of a circular orbit due to a central mass $M$ in Newtonian gravity
then
\beq
   \tilde F_{\rm(drag)} = \frac{A}{r^2} \left(\frac{M}{r}\right)^{1/2} = AM r^{-5/2} \,.
\eeq
The Newtonian gravitational force per unit mass $\tilde F_{\rm(g)}=M/r^2 $
only grows like the inverse square of the distance, so as one approaches the central mass, the drag force becomes more and more important compared to the gravitational free fall behavior of the particle initially in circular motion.
Similarly the radial radiation pressure force per unit mass under these conditions is just $A/r^2$ and the ratio of the drag force to the radial pressure force is just $\nu\ll1$, namely very small. This is the approach taken by Robertson in describing the radiation using special relativity and gravitation using Newtonian theory.

Since we use geometrical units  $c=1=G$,
to compare these formulas with the literature, one must restore the standard units. If $m$ is the mass of the particle under consideration, then the tangential drag force is actually
\beq
m\tilde F_{\rm(drag)} = \frac{W}{c^2} \nu
\eeq
where $\nu = (GM/r)^{1/2}$ is the Keplerian speed and $W= c\mathcal{L} \, \sigma /(4\pi r^2)$ is the power of the incoming radiation at the particle position expressed in terms of the luminosity $\mathcal{L}$ of the source and the cross-section $\sigma$ of the particle.
Introducing the Eddington luminosity
\beq
\mathcal{L}_{\rm Edd} 
= \frac{4\pi GM c m_p}{\sigma_{\rm T}} 
= 1.3\times 10^{38} \left(\frac{M}{M_\odot}\right) \, {\rm erg/s} \,, 
\eeq
where $m_p$ is the proton mass, $\sigma_{\rm T}$ is the Thompson cross-section and $M_\odot$ is the mass of the sun,
the drag force can be rewritten
\beq
  m\tilde F_{\rm(drag)}  
  = m \left( \frac{\mathcal{L} }{\mathcal{L}_{\rm Edd} } \right)
    \left( \frac{m_p }{m } \right)  \left( \frac{\sigma }{\sigma_{\rm T}} \right) \frac{(GM)^{3/2}}{r^{5/2}} \,.
\eeq
This identifies our radiation constant as
\beq
   A =  GM \left( \frac{\mathcal{L} }{\mathcal{L}_{\rm Edd} } \right)   
           \left( \frac{\sigma }{\sigma_{\rm T}} \right) 
           \left( \frac{m_p }{m } \right) \,.
\eeq
If we consider a hydrogen atom as representative of the particle, then its mass is $m_p$ and $\sigma_{\rm T}=\sigma$ leading to the ratio
\beq
   \frac{A}{GM} =   \left( \frac{\mathcal{L} }{\mathcal{L}_{\rm Edd} } \right)  
\eeq 
which is small for sources of astrophysical interest.

\subsection{Schwarzschild spacetime}

The Schwarzschild spacetime is characterized by the metric functions
\beq
\fl
g_{tt}=-N^2\,,\quad 
g_{t\phi}=0\,,\quad 
g_{rr}=1/N^2\,,\quad 
g_{\theta\theta}=r^2\,,\quad 
g_{\phi\phi}=r^2\sin^2\theta\,,
\eeq
where the lapse function is
\beq
N=\sqrt{1-\frac{2M}{r}}\sim 1-\frac{M}{r}\,,
\eeq
in which the approximate expression represents the asymptotic value at $r \to\infty$ and to first order in $M$.
In this case as well the ZAMOs are aligned with the coordinate time world lines; however, they form an accelerated ($a(n)^{\hat r}=M/(r\sqrt{r^2-2Mr})\sim M/r^2$) and  expansionfree ($\theta(n)=0$) congruence.
The Lie curvature of the $\phi$ loops has only a radial component with value 
$k_{\rm (lie)}(n)^{\hat r}=- N/r\sim -1/r+ M/r^2$.
Radially outgoing (geodesic) photons on the equatorial plane have 4-momentum
\beq
k = E\left[\left(1-\frac{2M}{r}\right)^{-1}\partial_t +\partial_r\right]
  = E(n)[n+e_{\hat r}]\,.
\eeq
Eqs.~(\ref{eqs:beta_eq_pi2_tau0}) reduce to
\begin{eqnarray}
\label{eqs:beta_eq_pi2_tau}
\frac{\rmd \nu}{\rmd \tau}
&=& -\frac{N\sin\alpha}{r\gamma}\nu_K^2 
   +\frac{A}{r^2N^2} (1-\nu \sin \alpha)(\sin \alpha -\nu)\,,\nonumber \\
\frac{\rmd \alpha}{\rmd \tau}
&=& \frac{N\gamma\cos \alpha }{r\nu}  (\nu^2-\nu_K^2) 
   +\frac{A}{r^2N^2\nu} (1-\nu \sin \alpha)\cos \alpha\,,
\nonumber\\
\frac{\rmd r}{\rmd \tau}
&=& \gamma \nu N \sin\alpha\,,
\end{eqnarray}
where we have used the relation $a(n)^{\hat r}=-k_{\rm (lie)}(n)^{\hat r} \nu_K^2$,
\beq
\nu_K = \sqrt{\frac{M}{r-2M}}
\eeq
is the Keplerian speed associated with circular geodesics.
If one is only interested in the spatial orbit of the particle,
one can choose $r$ or $\phi$ as the parameter along its path, re-expressing the above equations using the chain rule.

For the case $\nu=0$ of a particle at rest with respect to the ZAMOs, these equations reduce to the single condition Eq.~(\ref{eq:radialbalance}) representing the balancing of the gravitational attraction and the radiation pressure at constant $r$ and $\phi$, namely
\beq
\frac{A}{M}= \left(1-\frac{2M}{r}\right)^{1/2} 
\quad\rightarrow\quad
r= r_{\rm(crit)} \equiv  \frac{2M}{1-A^2/M^2}
\,.
\eeq
This is exactly the zero speed singular point discussed for radial motion taking into account the finite radius of the photon source by Abramowicz et al \cite{abr-ell-lan} (which therefore introduces a complicating geometric factor into the problem incorporating the effective solid angle of photons which arrive from the entire surface of the photon source) and makes the general relativistic case quite different from the Robertson limit where the balance condition is independent of radius. Instead for the general relativistic case in the present context, if $A/M>1$, the net effect is always expulsion from the central object as in the Robertson limit, but if $A/M<1$, there is always exactly one radius $r_b\in (2M,\infty)$ at which force balance is achieved for zero speed, and this point in the system of differential equations behaves like an attractor for some subspace of the initial data.
If $A/M\ll 1$, then this radius is near $r=2M$ and hence not important for a real star. In fact in the Newtonian limit this condition reduces to simply $A/M=1$ which is independent of the radius.

This equilibrium condition can also be re-expressed in terms of the Eddington luminosity as above using its general relativistic generalization \cite{abr-ell-lan}
\beq 
\mathcal{L}_{\rm EddGR} = \mathcal{L}_{\rm Edd}/(1-2M/R)^{1/2} 
\,, 
\eeq
where $R$ is the radius of the emitting body. Then the equilibrium condition is
\beq
   \frac{A}{GM} =   \left( \frac{\mathcal{L} }{\mathcal{L}_{\rm EddGR} } \right) 
                   \left(\frac{1-{2M}/{r}}{1-{2M}/{R}}\right)^{1/2} 
           \left( \frac{\sigma }{\sigma_{\rm T}} \right) 
           \left( \frac{m_p }{m } \right)
\,,
\eeq 
so that when evaluated for a proton at the surface of the star where the Thompson cross-section is relevant, one has a balance of the outward radiation pressure and the inward gravitational force.

The system of differential equations (\ref{eqs:beta_eq_pi2_tau}) requires numerical solution, easily done with a computer algebra system (Maple was used here).
One can study the initial behavior for a solution which starts out close to a circular geodesic at a fixed radius $r_0$ and constant speed $\nu_K$ to see how the orbit begins to decay for the more interesting case $A/M<1$. This is done in appendix B.

Figs.~1 and 2 show some typical solution curves starting from initially circular initial data either inside (Fig.~1) or outside (Fig.~2) the critical radius at which a particle at rest with respect to the ZAMOs (which in turn are at rest with respect to the coordinate system) remains at rest. For comparison, geodesics with the same initial data are shown in gray.
Unless the particle has a sufficiently high initial speed that it can escape to infinity, it is  forced to migrate to this critical radius where it comes to rest. Fig.~2 shows two final cases where the initial speed is sufficient for either the solution curve or the geodesic to escape. 
If the initial radius is taken far outside the critical radius, the circular orbit gradually decays as in the Robertson limit until it finishes at the critical radius.

\begin{figure} 
\typeout{*** EPS figure 1}
\begin{center}
\includegraphics[scale=0.3]{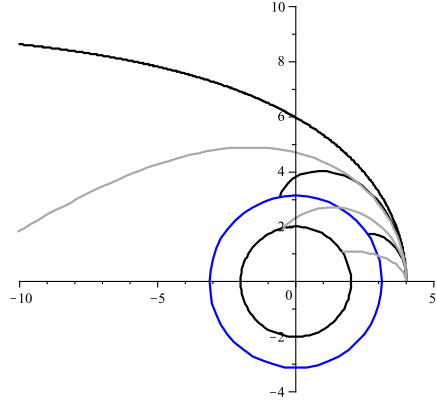}
\end{center}
\caption{The orbit of the particle in the Schwarzschild spacetime with $M=1$, $A/M=0.6$, $r_{\rm(crit)}=3.125M$, $\nu_K=0.7071$.
The inner circle is the horizon $r=2M$, while the outer circle is at the critical radius which is inside the initial data position. 
Initial conditions have $(r(0),\phi(0),\alpha(0))=(4M,0,0)$
and $\nu(0)=0.2,0.5,0.8$. The corresponding geodesics $A/M=0$ are in gray.
} 
\label{fig:1}
\end{figure}

\begin{figure} 
\typeout{*** EPS figure 2}
\begin{center}
\includegraphics[scale=0.3]{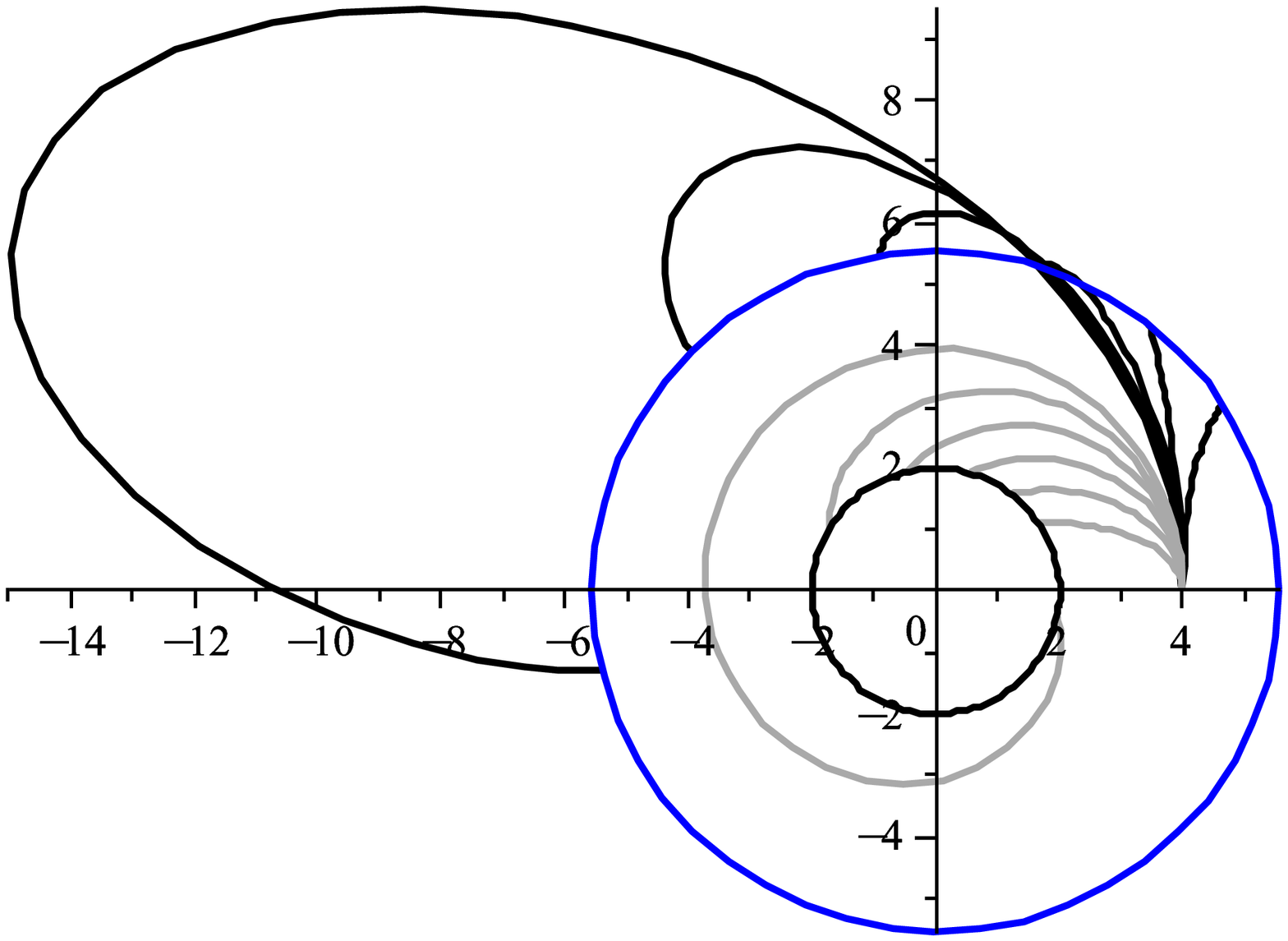}
\includegraphics[scale=0.35]{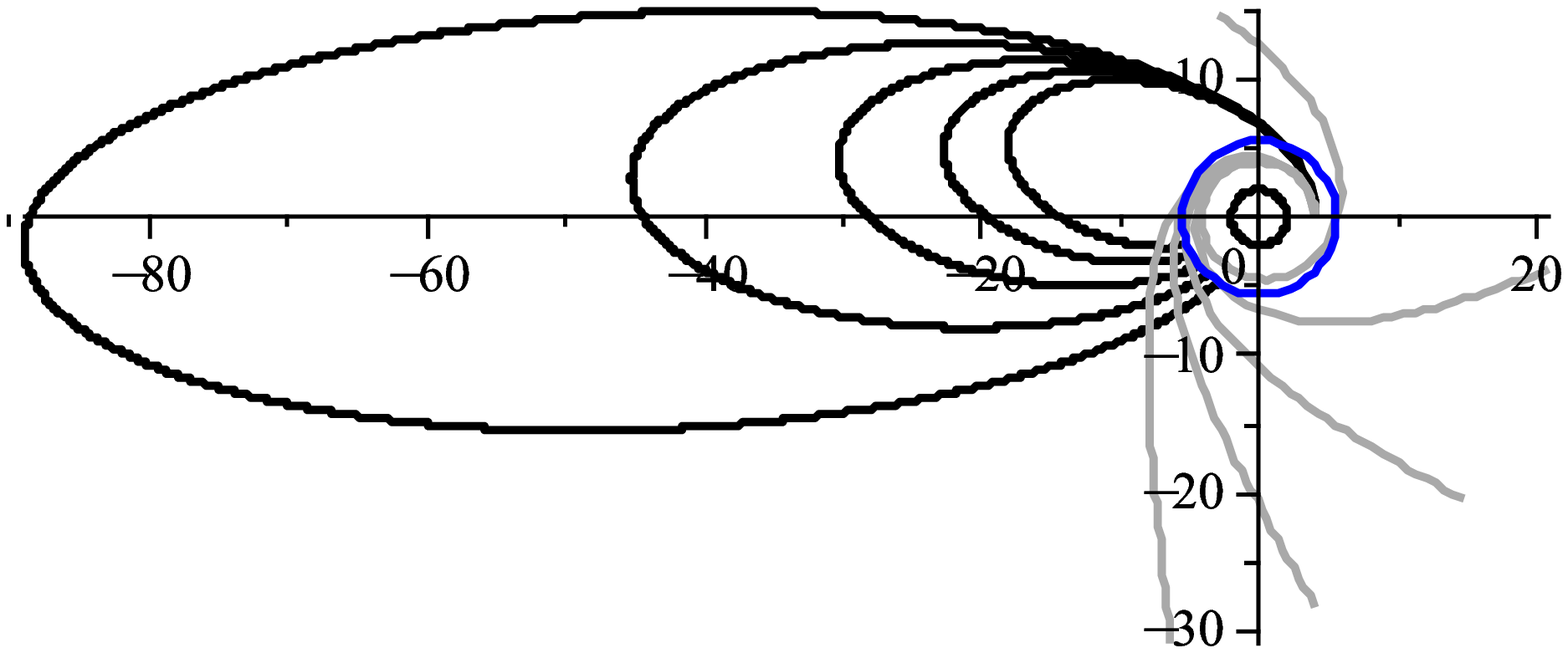}
\end{center}
\caption{The orbit of the particle in the Schwarzschild spacetime with $M=1$, $A/M=0.8$, $r_{\rm(crit)}=5.5M$, $\nu_K=0.7071$. 
The inner circle is the horizon $r=2M$, while the outer circle is at the critical radius which is outside the initial data position.
Initial conditions have $(r(0),\phi(0),\alpha(0))=(4M,0,0)$ and for the left figure
$\nu(0)=0.2,0.3,\ldots,0.7<\nu_K$ while for the right figure $0.71,0.72,\ldots,0.75>\nu_K$. 
The corresponding geodesics $A/M=0$ are in gray.
} 
\label{fig:2}
\end{figure}

\begin{figure} 
\typeout{*** EPS figure 3}
\begin{center}
\includegraphics[scale=0.3]{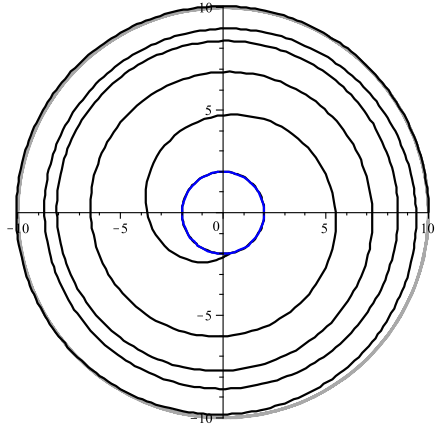}
\end{center}
\caption{The orbit of the particle in the Schwarzschild spacetime with $M=1$, $A/M=0.01$, $r_{\rm(crit)}\approx 2M$ and initial conditions
$(r(0),\phi(0),\alpha(0))=(10M,0,0)$
with the circular geodesic speed $\nu(0)=\nu_K=0.3536$. The circular geodesic (gray) is shown with the in-spiraling orbit (black) with the same initial conditions.
} 
\label{fig:3}
\end{figure}

\subsection{Kerr spacetime}

In the equatorial plane of the Kerr metric, the metric is
\begin{eqnarray}
\fl\qquad
&& g_{tt}=-\left(1-\frac{2M}{r}\right)\,,\quad 
g_{t\phi}=-\frac{2aM}{r}\,, \quad 
g_{\phi\phi}=\frac{r^3+a^2r+2a^2M}{r}\,,\nonumber \\
\fl\qquad
&& g_{rr}=\frac{r^2}{\Delta}\,,\quad g_{\theta\theta}=r^2\,,
\end{eqnarray}
so that
\beq
\fl
N=\sqrt{\frac{r\Delta}{r^3+a^2r+2a^2M}}\sim 1-\frac{M}{r}\,,\qquad
N^\phi=-\frac{2aM}{r^3+a^2r+2a^2M}\sim -\frac{2aM}{r^3}\,,
\eeq
where $\Delta=r^2+a^2-2Mr$ and the approximate expressions represent their asymptotic values ($r\to \infty$) to first order in $M$. The ZAMOs are timelike outside the horizon $r_+=M+\sqrt{M^2-a^2}$.
The nonvanishing components of the ZAMO kinematical fields are
\begin{eqnarray}
a(n)^{\hat r}
&=\frac{M[(r^2+a^2)^2-4a^2Mr]}{r^2\sqrt{\Delta}(r^3+a^2r+2a^2M)}&\sim \frac{M}{r^2}\,,\nonumber \\
\theta_{\hat \phi}(n)^{\hat r}
&=-\frac{aM(3r^2+a^2)}{r^2(r^3+a^2r+2a^2M)}&\sim -\frac{3aM}{r^3}\,,\nonumber \\
k_{\rm (lie)}(n)^{\hat r}
&=-\frac{\sqrt{\Delta}(r^3-a^2M)}{r^2(r^3+a^2r+2a^2M)}&\sim -\frac{1}{r}+\frac{M}{r^2}\,.
\end{eqnarray}
Circular geodesics correspond to orbits
\beq
\fl\qquad
U_\pm=\gamma_\pm (n+\nu_\pm e_{\hat \phi}), \qquad \nu_\pm =\frac{a^2\mp 2a\sqrt{Mr}+r^2}{\sqrt{\Delta}(a\pm r\sqrt{r/M})},
\eeq
and the following relation between $\nu_\pm$ and the ZAMO kinematical fields hold
\beq
a(n)^{\hat r} = k_{\rm (lie)}(n)^{\hat r} \nu_+\nu_- \,, \quad 
-2\theta_{\hat \phi}(n)^{\hat r}
= k_{\rm (lie)}(n)^{\hat r}( \nu_+ + \nu_-).
\eeq

Equations (\ref{eqs:beta_eq_pi2_tau0}) can then be rewritten as
\begin{eqnarray}
\label{eqs:beta_eq_pi2_tau0_kerr}
\fl
\frac{\rmd \nu}{\rmd \tau}
&=& 
-\frac{\sin\alpha k_{\rm (lie)}(n)^{\hat r}}{\gamma}[\nu_+\nu_-  -\nu\cos \alpha\, (\nu_++\nu_-)]
  +\frac{A(1-\nu \sin \alpha)(\sin \alpha -\nu)}{r\sqrt{g_{\phi\phi}}N^2} \,,
\nonumber \\
\fl
\frac{\rmd \alpha}{\rmd \tau}
&=& 
 -\frac{\gamma\cos \alpha k_{\rm (lie)}(n)^{\hat r}}{\nu}
   [\nu_+\nu_- -\nu\cos \alpha\, (\nu_++\nu_-) +\nu^2 ] 
+ \frac{A(1-\nu \sin \alpha)\cos \alpha}{r\sqrt{g_{\phi\phi}}N^2\nu} \,,
\nonumber\\
\fl
\frac{\rmd r}{\rmd \tau}
&=& 
\frac{\gamma \nu \sin\alpha}{\sqrt{g_{rr}}}
\,.
\end{eqnarray}

For the case $\nu=0$ of a particle remaining at rest with respect to the ZAMOs, these reduce to the single radial force balance condition
\beq
\frac{A}{M}=\frac{[(r^2+a^2)^2-4a^2Mr]\sqrt{\Delta}}{r^2 (g_{\phi\phi})^{3/2}}\,,
\eeq
which cannot be solved explicitly for $r$. However,
the right hand side of this equation takes values between 0 at the horizon $r=r_+$ and 1 when as $r\to\infty$ so a critical radius $r_{\rm(crit)}$
always exists for which this is satisfied for any proper fractional value of $A/M$. If $A/M>1$ of course no static solutions exist.
This balance condition can also be used to generalize the expression for the Eddington luminosity to this case and rewrite the balance condition in terms of it as done for the Schwarzschild case.

\begin{figure} 
\typeout{*** EPS figure 4}
\begin{center}
\includegraphics[scale=0.3]{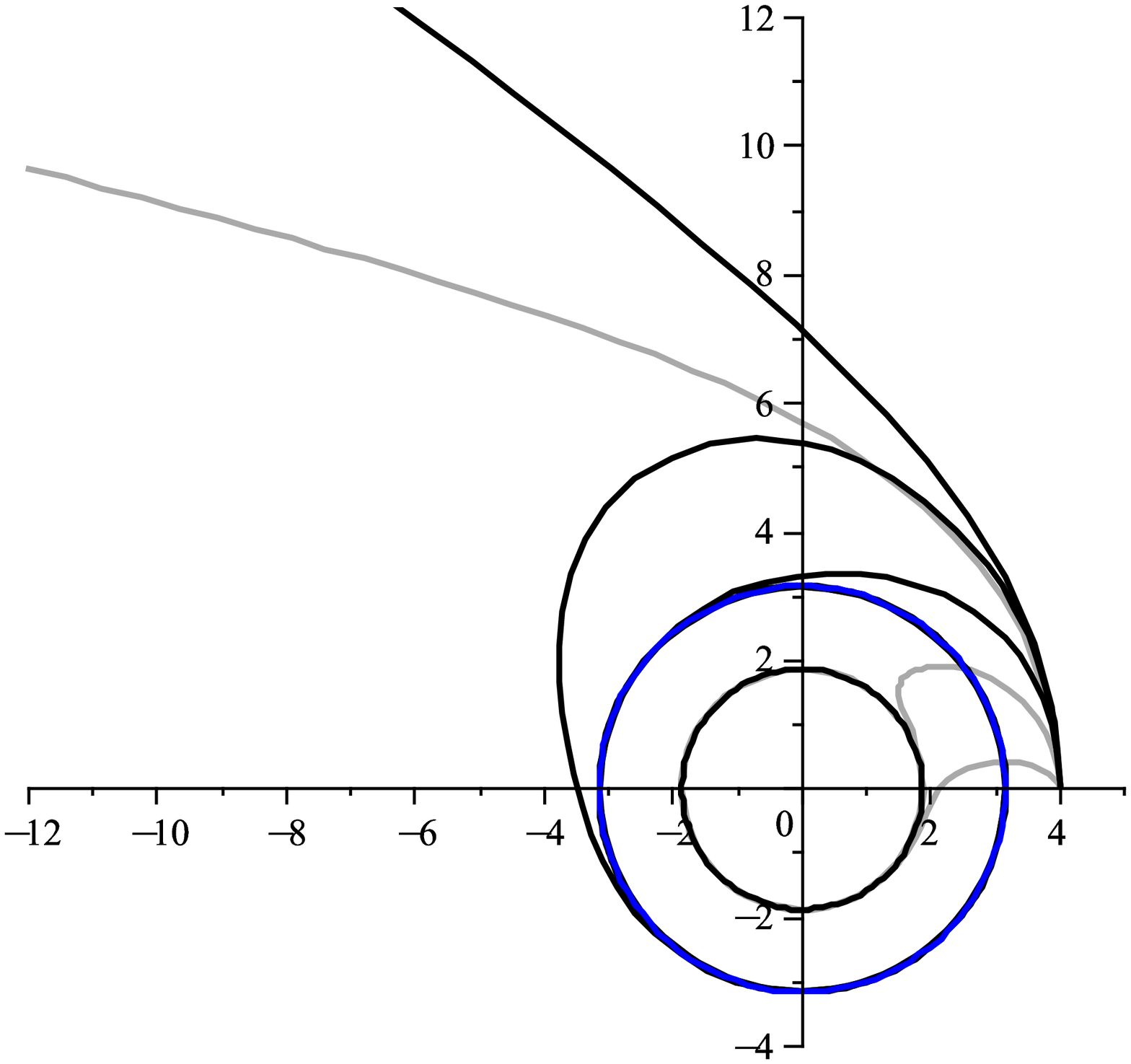}
\includegraphics[scale=0.3]{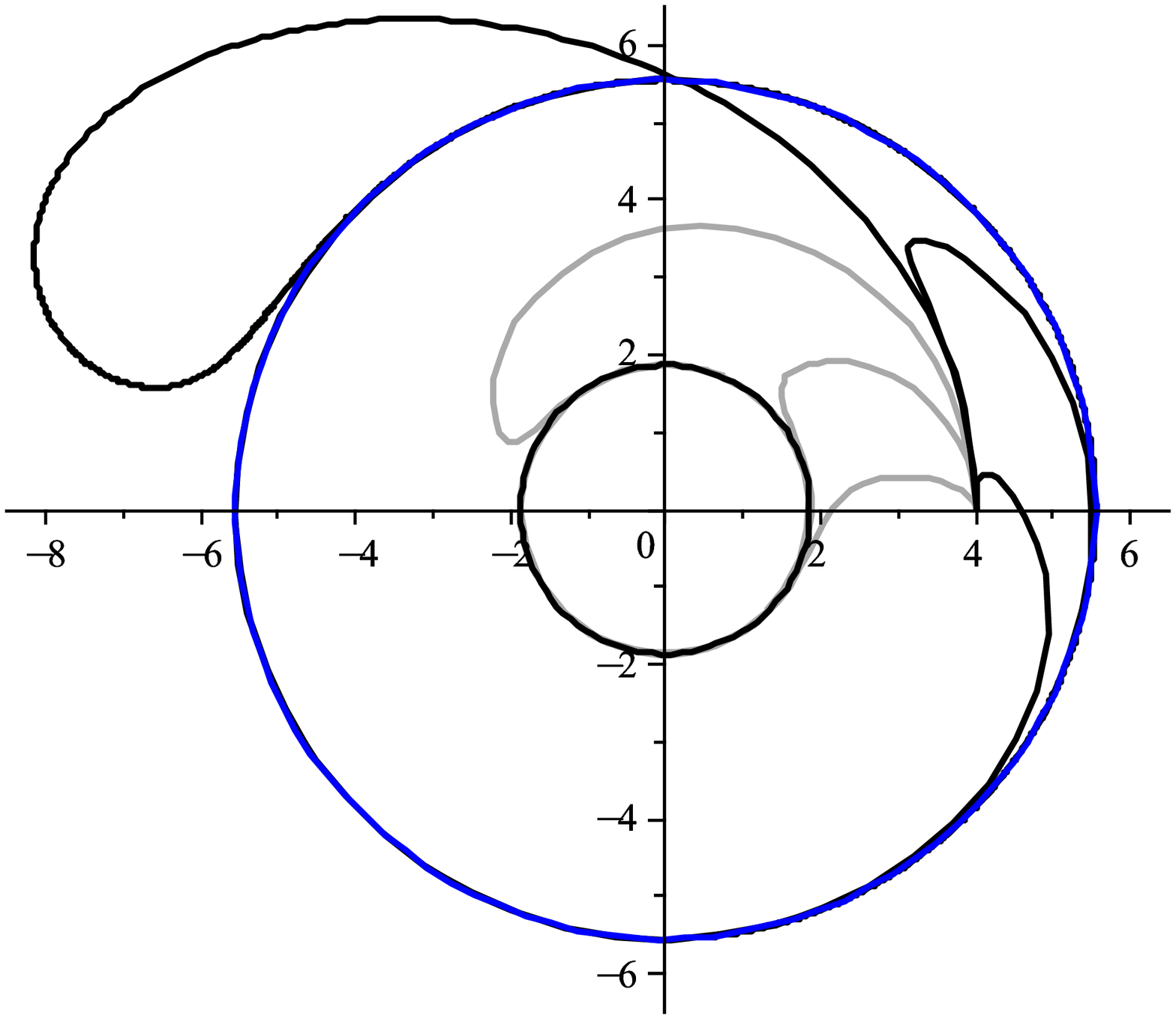}
\end{center}
\caption{The orbit of the particle in the Kerr spacetime with $M=1$, $a=0.5$ (left figure), $a=-0.5$ (right figure), $A/M=0.6$, $r_{\rm(crit)}=3.154M$. 
The inner circle is the horizon $r=1.866 M$, while the outer circle is at the critical radius which is inside the initial data position.
Initial conditions have $(r(0),\phi(0),\alpha(0))=(4M,0,0)$
and $\nu(0)=0.2,0.5,0.8$. The corresponding geodesics $A/M=0$ are in gray.
The bound orbits end up co-rotating with the hole at the horizon (geodesics) or at the critical radius (accelerated).} 
\label{fig:4}
\end{figure}

\begin{figure} 
\typeout{*** EPS figure 5}
\begin{center}
\includegraphics[scale=0.3]{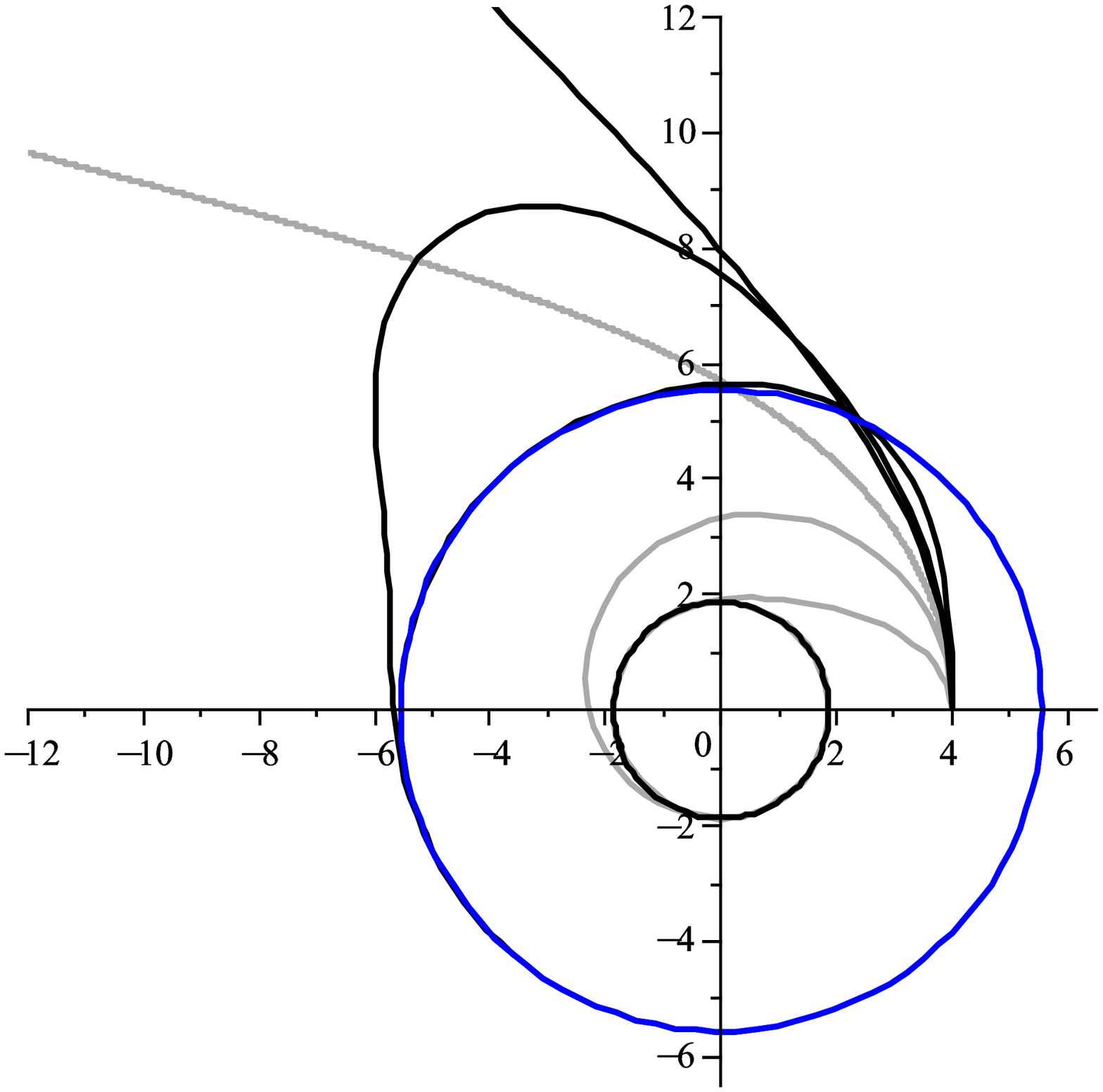}
\includegraphics[scale=0.3]{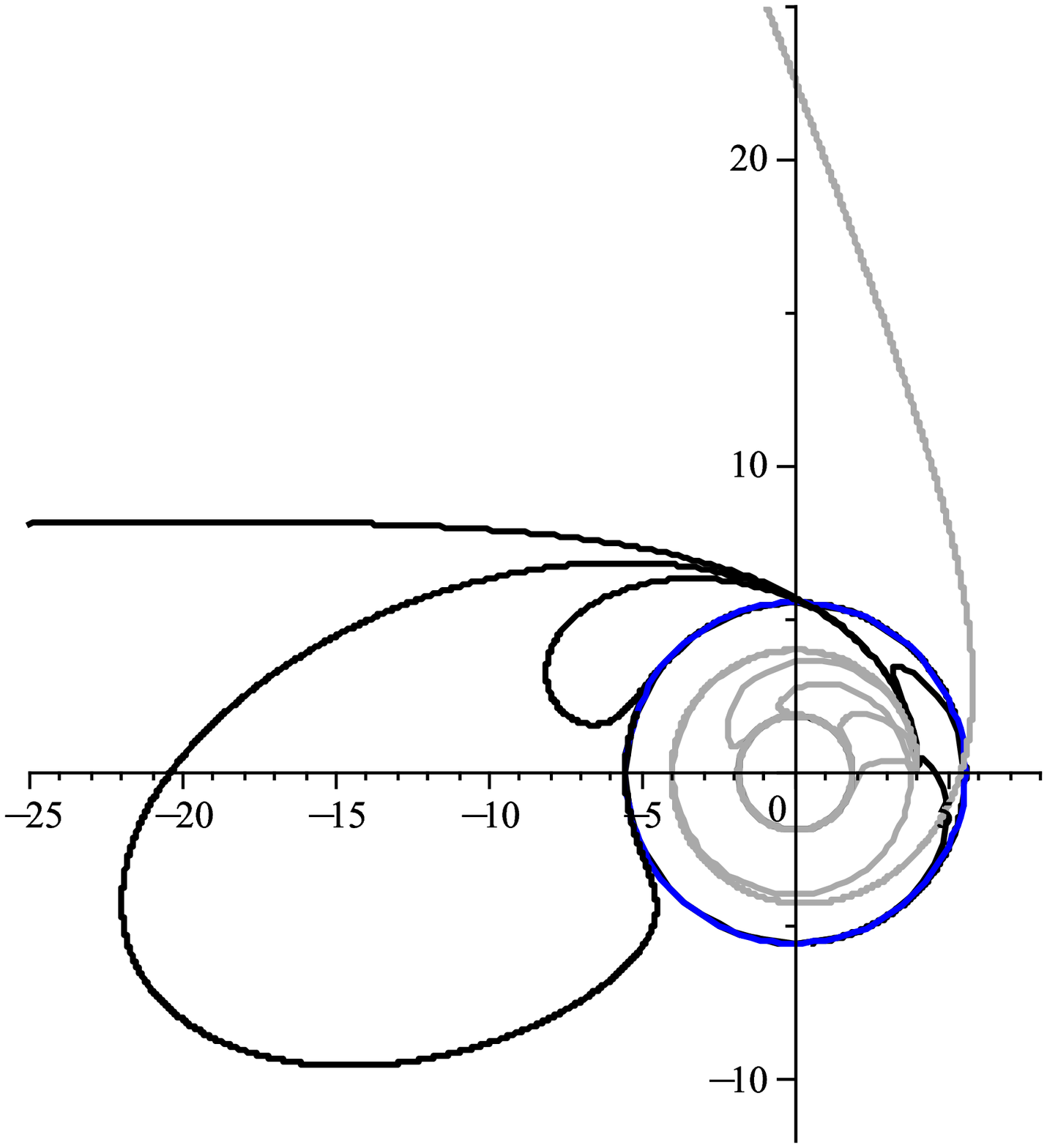}
\end{center}
\caption{The orbit of the particle in the Kerr spacetime with $M=1$, $a=0.5$ (left figure), $a=-0.5$ (right figure), $A/M=0.8$, $r_{\rm(crit)}=5.551M$. 
The inner circle is the horizon $r=1.866 M$, while the outer circle is at the critical radius which is outside the initial data position.
Initial conditions have $(r(0),\phi(0),\alpha(0))=(4M,0,0)$
and $\nu(0)=0.2,0.5,0.8$ for the left figure, while in the right figure $\nu(0)=0.2,0.5,0.8,0.847$ for both the accelerated and geodesic curves and then finally $\nu(0)=0.9$ for the accelerated curve and $\nu(0)=0.848$ for the geodesic, both of which escape to infinity.
The corresponding geodesics $A/M=0$ are in gray.
The bound orbits end up co-rotating with the hole at the horizon (geodesics) or at the critical radius (accelerated).
} 
\label{fig:5}
\end{figure}

Fig.~3  shows some typical solution curves starting from initially circular initial data outside the critical radius at which a particle is at rest with respect to the ZAMOs in the case of positive rotation parameter $a=0.5$ corresponding to counterclockwise dragging (left) and in the case of negative rotation parameter $a=-0.5$ corresponding to clockwise dragging (right). The same qualitative orbit behavior results as in the Schwarzschild spacetime but here the ZAMOs themselves are dragged along by the spacetime geometry, so that the actual curves in the plane are severely distorted by this effect. Unless the initial velocity is sufficiently high, the solution curves end up at corotating with the geometry at the critical radius. Fig.~4 shows the same situation for the case in which initial data is taken inside the critical radius. Initial data is taken at rather small radii not far from the horizon only to show the effects of the radiation pressure in their most exaggerated form, and is not intended to model actual astrophysical scenarios.

\section{Concluding remarks}

It is remarkable that in the seventy years since Robertson studied this effect, there seems to be no published article which reconsiders it within the full theory of general relativity. It is a clean geometric model that is a simple extension of geodesic motion and with today's computer algebra systems, one can plot orbits accurately within a matter of seconds to explore its consequences numerically. Because Robertson was himself repeating earlier nonrelativistic calculations, his own article omitted many details which we have provided here in the framework of stationary axisymmetric spacetimes, made explicit for the Schwarzschild and Kerr spacetimes. 
The model we developed here, while still a toy model, can
provide interesting applications to astrophysical
problems, especially in relation to accreting black holes and
neutron stars. The inclusion of effects such as non-equatorial
particle orbits and photon fields endowed with angular momentum
will be especially relevant in this context.

\appendix
\section{Weak field, slow motion, small drag limit}

Robertson developed the relativistic equations appropriate to describe this problem in special relativity and then looked at the Newtonian approximation, taking into account by hand the perihelion precession due to general relativity. His results can be obtained from the Schwarzschild case in the weak field, slow motion approximation for a small drag coefficient. Letting $\dot f= df/dt$, and introducing the approximation
$$
   t \to \tau \,,\quad
   \left| \frac{dr}{dt}\right| \ll 1\,,\quad
  \frac{M}{r} \ll 1\,,\quad
  \frac{A}{r} \ll 1\,,
$$
one obtains
$$
\ddot r - r \dot\phi^2 = -\frac{(M-A)}{r^2} -3M \dot\phi^2 -\frac{2A \dot r}{r^2} \,,\quad
(r^2\dot\phi)\,\dot{} =-A\dot\phi\,.
$$
The azimuthal equation leads to the constant of the motion
$$
  h = r^2\dot\phi+A\phi
$$
so that $H\equiv r^2\dot\phi=h-A\phi$.
Backsubstituting this into the radial equation, introducing the dimensionless reciprocal variable $u=M/r$, and changing the independent variable from $t$ to $\phi$ using $d\phi/dt = H/r^2$, one finds
$$
  \frac{d^2 u}{d\phi^2}+ u = \frac{M(M-A)}{H^2}  -\frac{A}{H}\frac{du}{d\phi} + 3 u^2\,.
$$
The first term on the right hand side represents the Newtonian gravitational force slightly reduced by the radiation pressure which leads to elliptical orbits, the last term the general relativistic term responsible for the perihelion precession of those elliptical orbits, and the middle term the drag effect responsible for the decay of the orbit. 

In the absence of the precession term this reduces to
\beq
\label{eq_app}
\frac{\rmd^2 u}{\rmd \phi^2}+\frac{A}{(h-A \phi)}\frac{\rmd u}{\rmd \phi }+u=\frac{M(M-A)}{(h-A \phi)^2}\,,
\eeq
which is further simplified as noted by Robertson by introducing the new variable $x$ such that $\phi=-x+h/A$
\beq
\frac{\rmd^2 u}{\rmd x^2}-\frac{1}{x}\frac{\rmd u}{\rmd x}+u = \frac{q}{x^2}\,,\quad 
q=\frac{M(M-A)}{A^2}
\eeq
which admits the following solution in the notation of Maple
\beq
\fl\quad
u=x[C_1\, {\rm BesselJ} \left( 1,x \right) + C_2\, {\rm BesselY} \left( 1,x \right)]
   +q\, x{\rm LommelS}_1 \left( -2,1,x \right)\,,
\eeq
with the consequence
\beq
\label{der_u}
\fl\quad
\frac{\rmd u}{\rmd x}
=x[C_1\, {\rm BesselJ} \left(0,x \right)+C_2\, {\rm BesselY} \left(0,x \right)]
   -2q\, x{\rm LommelS}_1 \left( -3,0,x \right)\,. 
\eeq
This exact solution $u(x)$ of the approximate equations can be used to express $\alpha(x)$ and in turn $\nu(x)$. In fact, using the approximate relations
\beq
\frac{\rmd r}{\rmd t}=\nu \sin \alpha\,,\qquad 
\frac{\rmd \phi}{\rmd t}=\frac{\nu}{r} \cos \alpha\,,
\eeq
one gets
\beq
\frac{\rmd r}{\rmd \phi}
=r\tan \alpha \quad \rightarrow \quad \cot\alpha = u \frac{\rmd u}{\rmd x}\,,
\eeq
where $\rmd u/\rmd x$ is given by Eq.~(\ref{der_u}).

However, one can see how the approximately elliptical orbits evolve including the relativistic precession term by using the further approximation of small drag $A/h \ll 1$, so that
\beq
  \frac{1}{H} \approx \frac{1}{h} \left( 1+\frac{A}{h}\phi \right) \,.
\eeq
Then radial equation linearized in this ratio leads to
\beq\label{eq:udeq2}
  \frac{d^2 u}{d\phi^2}+ u 
= \frac{M(M-A)}{h^2} \left( 1+\frac{2A}{h}\phi \right)
 + 3 u^2 -\frac{A}{h}\frac{du}{d\phi}\,.
\eeq
Solving this in the absence of the $A/h$ and precession correction terms leads to a Newtonian elliptical orbit
\beq
   u_{(0)} =\frac{M}{p} (1 + e \cos\phi) \,,
\eeq
where the orbital parameters correspond to an adjusted central force $M\to M-A$
\beq
   p =a(1-e^2) = \frac{h^2}{(M-A)}
\,. 
\eeq

Excluding the single nonlinear precession term and using an osculating ellipse approach to investigate the secular perturbations of these elliptical orbits to first order in $A/h$ and to first order in $u\ll 1$, one looks for approximate next order solutions 
$$
u(\phi)
  = u_{(0)}(\phi)+u_{(1)}(\phi)
$$
and finds
$$
u_{(1)}(\phi) =  \frac{M}{ph} \left(2\phi -\frac12 e\phi\cos\phi  \right) 
$$
Recombining the various terms to first order in  $A/h$ leads to
$$
 u(\phi) = \frac{M}{p(\phi)} (1 + e(\phi) \cos(\phi-\delta(\phi)) \,. 
$$
where
$$
  p(\phi) = p\left(1-2\frac{A}{h}\phi \right)\,,\quad
  e(\phi) = e\left(1-\frac{5}{2}\frac{A}{h}\phi \right)\,,\quad
  \delta(\phi) = 0\,,
$$
which correspond to the following differential equations satisfied by the orbital elements for their logarithmic rates of change and in turn that of $a=p/(1-e^2)$ using logarithmic differentiation
$$
  \frac{d\ln p}{d\phi} = -2\frac{A}{h}\,,\
  \frac{d\ln e}{d\phi} = -\frac{5}{2}\frac{A}{h}\,,\
  \frac{d\ln a}{d\phi} = -\frac{2+3e^2}{1-e^2} \frac{A}{h}\
\,.
$$
To convert the $\phi$ derivative into a $t$ derivative in these equations, one can take the average related rate over one period of the zeroth order solution
$$
  \frac{dt}{d\phi} 
   = \frac{r^2}{H}
   = \frac{M^2}{H u^2}  \to 
  \frac{dt_{(0)}}{d\phi} 
   = \frac{M^2}{h u_{(0)}^2}
   = \frac{p^2}{h(1+e\cos\phi)^2}
   = \frac{[a(1-e^2)]^{2}}{h(1+e\cos\phi)^2}
$$
by calculating
$$
  \left\langle \frac{dt}{d\phi} \right\rangle
 = \frac{ \Delta t|_{\Delta\phi=2\pi} }{2\pi} 
 = \frac{[a(1-e^2)]^{2}}{2\pi h} \int_0^{2\pi} \frac{d\phi}{(1+e^2\cos\phi)^2} 
 = \frac{a^{2}(1-e^2)^{1/2}}{h}
$$
from which Robertson's results follow in the form similar to that given by Wyatt and Whipple \cite{wyatt} for the secular changes in these orbital parameters
$$
  \frac{d\ln p}{dt} = -2\frac{A}{a^{2}}\,,\
  \frac{d\ln e}{dt} = -\frac{5}{2}\frac{A}{a^{2}}\,,\
  \frac{d\ln a}{dt} = -\frac{2+3e^2}{1-e^2} \frac{A}{a^{2}}
\,.
$$

The quadratic precession term can then be handled again by considering a second linearized term $u_{(2)}(\phi)$ in the solution, following the discussion of Ohanian and Ruffini \cite{ruffohanian} 
$$
u(\phi)
  = u_{(0)}(\phi)+u_{(1)}(\phi) + u_{(2)}(\phi) \,.
$$
Expanding the $3u^2$ term about the zeroth order solution $3u_{(0)}^2$ in the original second order differential equation (\ref{eq:udeq2}), one  first neglects the small $u_{(1)}$ and $u_{(2)}$ correction terms in the square compared to $3u_{(0)}^2$, leading to
\begin{eqnarray}
\fl\qquad
\frac{d^2 u_{(2)}}{d\phi^2}&&\kern-13pt
  + u_{(2)}
 = 3u_{(0)}^2
\nonumber\\
\fl\qquad
 &=& 3\frac{M^2}{p^2} (1+ 2e\cos\phi +e^2\cos^2\phi)
\nonumber\\
\fl\qquad
 &=& 3\frac{M^2}{p^2} \left( 1+\frac12 e^2 + 2e\cos\phi +\frac12 e^2\cos 2\phi \right)
\,.
\end{eqnarray}
The $\cos 2\phi$ term on the right hand side leads to small oscillations in the solution for $u_{(2)}$, and the constant term a slight constant shift, but the $\cos\phi$ term is in resonance with the left hand side oscillator and will dominate both over the long term, leading to the secular part of the solution
\beq
    u_{(2)}(\phi) = 3\frac{M^2 e}{p^2} \phi\sin\phi \,.
\eeq
Reconstructing $u$ to first order in this small quantity then leads to the same form as above but with the additional precession term
\beq
  \delta(\phi) = 3\frac{M}{p} \phi \,,
\eeq
which agrees with Robertson's result when $A\ll M$.

\section{Perturbing initially circular orbits in Schwarzschild spacetime}

To consider an orbit in the Schwarzschild spacetime close to a circular geodesic 
\beq
\alpha_0(\tau)=0\,,\quad 
\nu(\tau)=\nu_K\,,\quad 
r(\tau)=r_0 \,.
\eeq
at a fixed radius $r_0$ and constant speed $\nu_K$ for the case of a small radiation pressure term whose effect will be small at least initially,
we linearize Eqs.~(\ref{eqs:beta_eq_pi2_tau}) with respect to the dimensionless parameter $\epsilon=A/M$ which is assumed to be much less than 1 by expanding those equations as follows
\beq
\fl\qquad
\alpha(\tau) = \epsilon\, \alpha_1(\tau)\,,\quad 
\nu(\tau) = \nu_K + \epsilon\, \nu_1(\tau)\,,\quad 
r(\tau) = r_0 + \epsilon\, r_1(\tau) \,.
\eeq

Inserting these quantities into Eqs.~(\ref{eqs:beta_eq_pi2_tau}) and linearizing them in $\epsilon$ we find
\begin{eqnarray}
\label{perturb}
\frac{\rmd \nu_1}{\rmd \tau}
&=& -\frac{\zeta_K\nu_K}{\gamma_K}\alpha_1-\frac{\nu_K^3}{r_0}\,,\nonumber \\
\frac{\rmd \alpha_1}{\rmd \tau}
&=& \frac{2\zeta_K\gamma_K}{\nu_K}\nu_1+\frac{\nu_K^2\zeta_K\gamma_K}{M}r_1+\frac{\nu_K}{r_0}\,,\nonumber \\
\frac{\rmd r_1}{\rmd \tau}&=& \gamma_K \zeta_K r_0 \alpha_1\,,
\end{eqnarray}
where $\zeta_K=(M/r_0^3)^{1/2}$ is the angular velocity associated with the circular geodesic.
We solve these with initial conditions representing the initial circular geodesic motion 
\beq
r_1(0)=0,\quad \alpha_1(0)=0,\quad  \nu_1(0)=0 \,, 
\eeq
leading immediately to
\beq
\frac{\rmd r_1}{\rmd \tau}(0) =0\,,
\quad 
\frac{\rmd \alpha_1}{\rmd \tau}(0) =\frac{\nu_K}{r_0}\,,
\quad  
\frac{\rmd \nu_1}{\rmd \tau}(0) =-\frac{\nu_K^3}{r_0} \,.
\eeq

Differentiating the second of Eqs.~(\ref{perturb}) and using the other two gives a decoupled equation for $\alpha_1$
\beq
\frac{\rmd^2 \alpha_1}{\rmd \tau^2}+\Omega^2 \alpha_1= b\,,
\eeq
where
\beq
  \Omega^2 = 
             \frac{ M (r_0-6M)}{r_0^3(r_0-3M)}\,, 
\qquad 
b = -\frac{2\zeta_K \gamma_K\nu_K^2}{r_0} \,.
\eeq
Note that $\Omega$ is a real frequency when $r_0>6M$
and coincides with the so called \lq\lq epicyclic frequency,"  
rescaled to correspond to a proper time parametrization of the orbit \cite{svmb}. 
The solution for $\alpha_1$ is given by
\beq
\alpha_1(\tau)=\frac{\nu_K}{\Omega^2 r_0}\left[\Omega\sin(\Omega \tau)+2\gamma_K\nu_K\zeta_K(\cos (\Omega \tau)-1)\right]\,,
\eeq
Introducing the notation
\beq
\fl
S(\tau)=\int^\tau_0 \alpha_1(\tau') \rmd \tau '=\frac{\nu_K}{\Omega^2r_0}\left[
(1-\cos(\Omega\tau))+2\nu_K\gamma_K\zeta_K\left(\frac{1}{\Omega}\sin(\Omega\tau)-\tau\right)\right].
\eeq

Then one immediately gets 
\beq
\fl\qquad
r_1(\tau) =\gamma_K\zeta_K r_0 S(\tau), \quad
\nu_1(\tau)=-\frac{\zeta_K\nu_K}{\gamma_K}S(\tau)-\frac{\nu_K^3}{r_0}\tau\,.
\eeq
Introducing the geodesic proper time orbital frequency
$\Omega_{\rm (orb)}=\gamma_K\nu_K/r_0$,    
the oscillation frequency for the perturbation can be expressed as
\beq
\frac{\Omega}{\Omega_{\rm (orb)}}= \sqrt{1-\frac{6M}{r_0}}\,.
\eeq
The final nonoscillatory term in $S(\tau)$ proportional to $\tau$ leads to the decrease of the radius causing the orbit to decay.

\end{document}